\documentclass{IEEEtran}
\usepackage{cite}
\usepackage{amsmath,amssymb,amsfonts}
\usepackage{graphicx}
\usepackage{textcomp,nicefrac}
\usepackage{subcaption}

\def\BibTeX{{\rm B\kern-.05em{\sc i\kern-.025em b}\kern-.08em
T\kern-.1667em\lower.7ex\hbox{E}\kern-.125emX}}
\begin{document}
\title{PCI-express based high-speed readout for the Belle~II DAQ upgrade}
\author{Q. D.~Zhou, S.~Yamada, P.~Robbe, D.~Charlet, R.~Itoh, \IEEEmembership{Senior Member, IEEE},  M.~Nakao, S.Y.~Suzuki, T.~Kunigo, E.~Jules, E.~Plaige, M.~Taurigna, H.~Purwar,  O.~Hartbrich,  M.~Bessner, K.~Nishimura, G.~Varner, Y. -T.~Lai,  T.~Higuchi, R.~Sugiura, D.~Biswas, and P. Kapusta
\thanks{This work was supported in part by Grants-in-Aid for Scientific Research -KAKENHI- JSPS International Research Fellow Grants No. 18F18324.}
\thanks{Q. D. Zhou is with Institute of Advanced Research and Kobayashi-Maskawa Institute, Nagoya Univ., Nagoya 464-8601, Japan (e-mail: qzhou@hepl.phys.nagoya-u.ac.jp).}
\thanks{S. Yamada, R. Itoh, M. Nakao, S.Y. Suzuki, and T. Kunigo are with
High Energy Accelerator Research Organization (KEK), Ibaraki 305-0801, Japan}
\thanks{P. Robbe is with the Univ. Paris-Saclay , CNRS/IN2P3, 
 the Laboratoire de Physique des Deux Infinis Irene Joliot-Curie (IJCLab), Orsay F-91898, France.}
\thanks{D. Charlet, E. Jules, E. Plaige, and M. Taurigna are with
 the Laboratoire de Physique des Deux Infinis Irene Joliot-Curie (IJCLab), Orsay F-91898, France.}
\thanks{ H. Purwar, O. Hartbrich, M. Bessner, K. Nishimura and G. Varner are with the Dept. of Phys. \& Astr., Univ. of Hawaii at M\=anoa, Honolulu, HI, 96822, USA.}
\thanks{ Y. -T. Lai, and T. Higuchi are with Kavli Institute for the Physics and Mathematics of the Universe, University of Tokyo, Kashiwa, 277-8583 Japan.}
\thanks{ R. Sugiura is with Tokyo University, Tokyo, 113-0033, Japan.}
\thanks{ D. Biswas is with University of Louisville, Louisville, Kentucky, 40292, USA.}
\thanks{ P. Kapusta is with Institute of Nuclear Physics (IFJ PAN), Krakow, 31-342, Poland.}
}

\maketitle

\begin{abstract}
Belle~II is a new-generation B-factory experiment, dedicated to exploring new physics beyond the standard model of elementary particles in the flavor sector. Belle~II started data-taking in April 2018, using a synchronous data acquisition (DAQ) system based on pipelined trigger flow control. The Belle~II DAQ system is designed to handle a 30-kHz trigger rate with approximately 1\% of dead time, under the assumption of a raw event size of 1 MB. The DAQ system is reliable, and the overall data-taking efficiency reached 84.2\% during the run period of January 2020 to June 2020. The current readout system cannot be operated in the term of 10 years from the viewpoint of DAQ maintainability; meanwhile, the readout system is obstructing high-speed data transmission. A solution involving a PCI-express-based readout module with high data throughput of up to 100 Gb/s was adopted to upgrade the Belle~II DAQ system. We particularly focused on the design of firmware and software based on this new generation of readout board, called PCIe40, with an Altera Arria 10 field-programmable gate array chip. Forty-eight GBT (GigaBit Transceiver) serial links, PCI-express hard IP-based DMA architecture, interface of timing and trigger distribution system, and slow control system were designed to integrate with the current Belle~II DAQ system. This paper describes the performances accomplished during the data readout and slow control tests conducted using a test bench and a demonstration performed using on-site front-end electronics, specifically involving Belle~II TOP and KLM sub-detectors. 
\end{abstract}

\section{Introduction}
\label{sec:introduction}
\IEEEPARstart{T}{he} standard model (SM) of particle physics, which describes the properties and physical behavior of elementary particles, has finally been established by the discovery of Higgs boson in 2012~\cite{Higgs_ATLAS, Higgs_CMS}. Although SM has demonstrated huge success, some phenomena are left unexplained, such as dark matter and material dominance universe. Belle~II \cite{belle2} is an experiment dedicated to exploring new physics beyond the SM in the flavor sector at the beam luminosity frontier. The SuperKEKB accelerator~\cite{superKEKB}, located at KEK, Tsukuba, Japan, is designed to use the ``nano-beam" scheme to achieve the instantaneous luminosity of $\rm{6.5} \times \rm{10}^{\rm{35}} $ $\mathrm{cm}^{-2}\mathrm{s}^{-1}$, which is approximately 35 times that of its predecessor KEKB accelerator, by colliding 7 GeV electrons and 4 GeV positrons. It can produce copious amounts of B and D mesons and $\tau$ leptons. The large statistics enables precise measurements of rare decays for testing SM with unprecedented sensitivity. The Belle~II experiment aims to collect 50 $ab^{-1}$ integrated luminosity by 2031. 

The Belle~II detector \cite{belle2} is a general-purpose spectrometer upgraded or replaced with respect to the Belle detector. It is configured around a 1.5 T superconducting solenoid surrounding the interaction region. The Belle~II detector consists of a vertex detector (VXD), central drift chamber (CDC), particle identification (PID) detector, electromagnetic calorimeter (ECL),  and neutral $K_{L}$ and Muon (KLM) detector, which are located from the inside to the outside of the detector. The VXD is made of a silicon pixel detector (PXD) and a silicon vertex detector (SVD) used for detecting decay vertices. The CDC located outside of the SVD is used as a tracking detector for precise measurement of momentum and energy loss of charged particles. Using the Cherenkov counter technique, the PID detector is divided into barrel and forward endcap parts, which are the time-of-propagation (TOP) counter and the proximity-focusing Aerogel Ring Imaging Cherenkov detector (ARICH), respectively. The ECL uses CsI(TI) crystals and CsI in the barrel and endcap regions, respectively. 
The KLM uses glass-electrode resistive plate chambers (RPCs) for the barrel region and layers of scintillator fibers for the endcap regions and the innermost layer of the barrel region.  

\section{Belle~II DAQ  and readout system}
\label{sec:DAQ}

\begin{figure}[t]
\centerline{\includegraphics[width=3.5in]{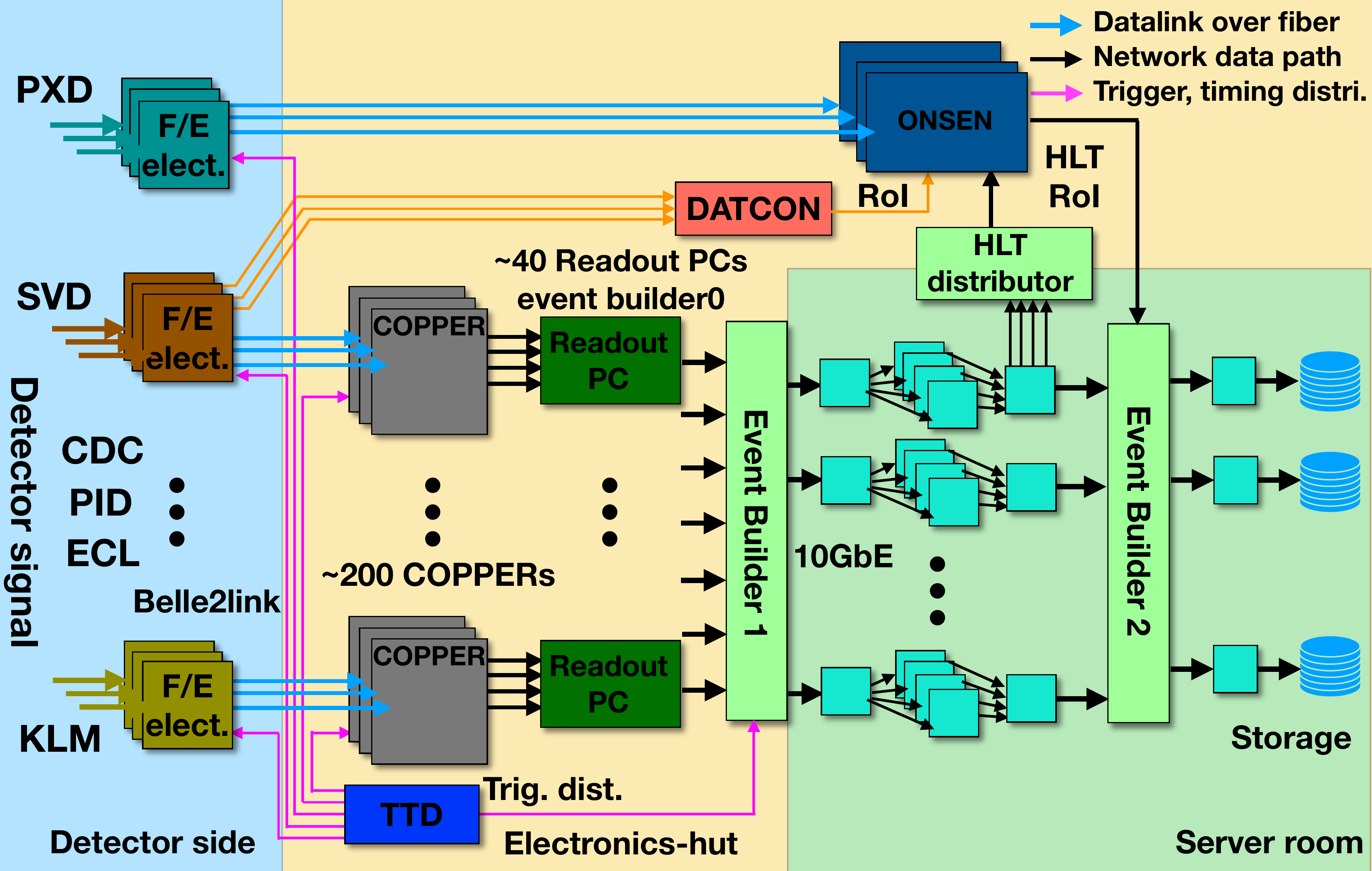}}
\caption{Block diagram of the current Belle~II DAQ system.}
\label{DAQ}
\end{figure}

\begin{figure}[t]
\centerline{\includegraphics[width=3.5in]{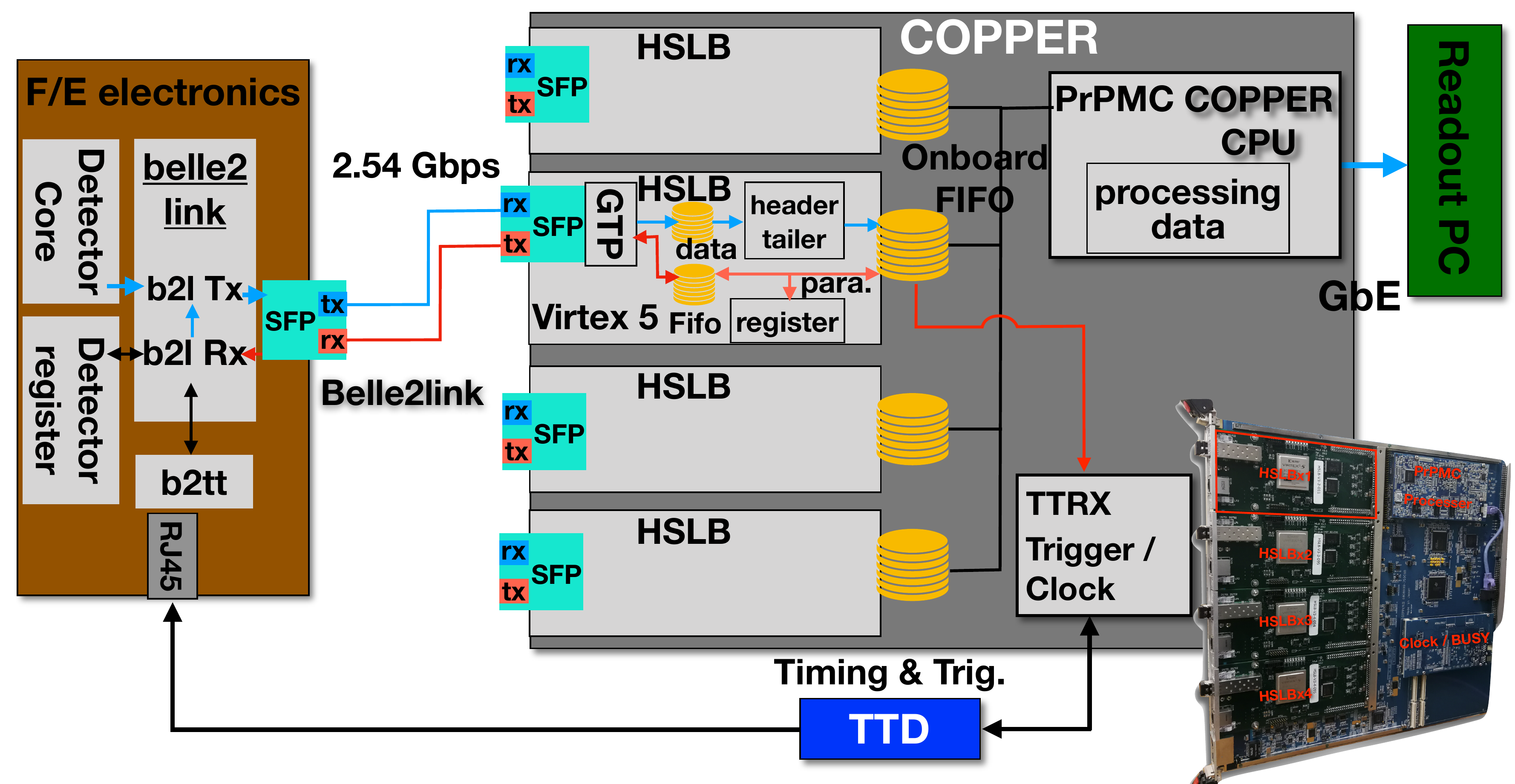}}
\caption{Schematic view of unified COPPER readout system.}
\label{copper}
\end{figure}

The Belle~II DAQ~\cite{b2daq,b2daq1} system is designed to process the data from the front-end electronics (FEE) to the storage system through several steps based on a highly unified system as shown in Fig.~\ref{DAQ}. The highly unified timing and trigger distribution (TTD) system~\cite{ttd1} first distributes Level 1 (L1) triggers~\cite{trigger} to each FEE board using Frontend Timing Switch (FTSW) modules~\cite{ttd} based on a 254 Mb/s serial link protocol called ``b2tt'' ~\cite{ttd}. The digitized signals from the FEE boards of each sub-detector, except for the PXD, are read out by a unified readout system based on the common module called Common Pipelined Platform for Electronics readout (COPPER)~\cite{copper}, as shown in Fig.~\ref{copper}. 
The data size of the PXD detector is too large to be read out by the COPPER module, an online event size reduction is necessary~\cite{performance}. 
A total of 203 COPPER boards are used in the current Belle~II DAQ system.
To unify the readout system, a high-speed data transmission based on a custom serial link protocol, called Belle2link~\cite{b2l}, is adopted. 
Xilinx GTP or GTX RocketIO up to 2.54 Gb/s, driven by the 127 MHz system clock, is used as the transceiver.
The required bandwidth of each data link is less than 1 Gb/s, while the throughput of each readout PC is also limited to 1 Gb/s. 
As shown in Fig.~\ref{copper}, each COPPER module hosts four custom High Speed Link Boards (HSLB). They are interconnected by a 32-bit peripheral component interconnect (PCI) bus and a local bus used for slow control with 7-bit and 8-bit widths for address and data, respectively.  
The data formatting (per link) and module-level event building are processed by an on-board CPU on the COPPER module; subsequently, the data from several COPPER modules are collected to a readout PC through the Gigabit ethernet. 
The event building and software event selection are performed by high-level trigger (HLT) farms~\cite{hlt}. Finally, the selected data are sent to the storage servers. To handle the large data flow at the beam intensity frontier, the Belle~II DAQ system is designed to handle 30 kHz of L1 trigger rate with approximately 1\% of dead time, under a raw event size of more than 1 MB.

\section{Requirements for Belle~II DAQ upgrade}
\label{sec:requirement}
The current readout system, mainly the COPPER module and its readout PC, is obstructing high-speed data transfer and DAQ maintainablility. The motivations for upgrading the Belle~II DAQ readout system are as follows:
\begin{enumerate}
\item As shown in Fig.~\ref{copper}, the COPPER module hosts four HSLB daughter cards used as single-link receivers, a processor PCI mezzanine card (PrPMC), and a custom daughter card used as the TTD interface. The broken parts are increasing, such as the chipset of PrPMC, because the COPPER module has been in use since the Belle experiment.
\item The data formatting and module-level event building are processed by an onboard CPU. The CPU usage has been observed to reach $\sim$ 60\% under the test condition of 1 kB/event/COPPER at a 30-kHz trigger rate. In addition, the smallest bandwidth of the current Belle~II DAQ system is the Gbit ethernet throughput of the COPPER module and the corresponding readout PC~\cite{performance}.
\item The current DAQ readout system could turn into a bottleneck when the Belle~II experiment takes an aggressive plan of operation in the next 10 years. For instance, the SuperKEKB accelerator can achieve higher luminosity than the designed value, the background level of the Belle~II experiment can be higher than expected, and the Belle~II experiment can implement a high rate of low multiplicity triggers and even realize a trigger-less DAQ. 
\end{enumerate}

To reduce the modifications required for the other systems, all the functionalities of the current COPPER-based readout system must be retained. For the new system and readout module, the following are required:
\begin{enumerate}
\item The Belle2link protocol needs to be continued so that no further modification is applied to the FEE's hardware and firmware for each sub-detector.  
\item The functionalities implemented in the COPPER module, including data handling, slow control, data-formatting for each link and each COPPER, event-building of merging data of 4 links into one event, and error checks, are essential for the new system.
\item To integrate into the TTD system, the TTD interface needs to be implemented on the new system. A busy handshake is needed for each link.  
\item The designed characteristics of the current readout system need to be followed. The trigger rate of 30 kHz with the estimated event size needs to be handled, 
onboard data corruption needs to be detected and should be reduced to less than 1 error~/~(100 kB/event × 30 kHz × 8 h). 
\end{enumerate}

Furthermore, the Belle~II experiment will run for at least 10 years, and an upgrade of any or all of its sub-detector is predictable. Thus, the new system should be capable of handling enhanced requirements from the sub-detector side. 
For instance, when the L1 trigger rate exceeds 30 kHz, the payload for each link will be higher than the bandwidth of COPPER system. 
In the Belle~II Collaboration, we had four proposals for the upgrade of the readout system, all of them had a well-established module used in other experiments. 
Three proposals have proceeded to the development of the firmware and software to verify that the boards satisfy the basic requirements of the Belle~II DAQ upgrade, as listed above. Finally, a PCI-express-based high-speed readout module, known as PCIe40~\cite{pcie40}, was adopted by the Belle~II Collaboration for DAQ upgrade. This is because PCIe40 fulfills the requirements of the Belle~II DAQ upgrade and a large amount of data can be transferred easily from a readout board to a PC server via PCI-express bus.  

\section{PCIe40 based new Belle~II readout system}
\label{sec:PCIe40}
PCIe40 is a new generation of readout modules based on the PCI-express Gen3 solution with a high data throughput of 100 Gb/s~\cite{pcie40_100G}.
This module will also be used for the DAQ upgrade of the LHCb  and ALICE experiments at the Large Hadronic Collider (LHC) \cite{pcie40, pcie40_ALICE}.  
To integrate with the current Belle~II DAQ system, we particularly focus on the design of the firmware and software for the PCIe40 board.
Since all functionalities of the current readout system are retained, 
no modification in the firmware and hardware of the sub-detector FEE is required. 
A single PCIe40 board can connect up to 48 bidirectional optical links, while only 4 bidirectional links can be handled by the COPPER module. 
Thus, the new readout system will become much more compact by replacing 203 COPPERs with 20 PCIe40 boards. The number of COPPERs used in the current readout system and the number of PCIe40 required for replacement are listed in Table~\ref{CPRvsPCIe40}. 

\begin{table}
\caption{Number of COPPER and PCIe40 required by the sub-detector}
\label{table}
\setlength{\tabcolsep}{3pt}
\begin{tabular}{|p{80pt}|p{70pt}|p{65pt}|}
\hline
Sub-detector&  Number of COPPER&  Number of PCIe40\\
\hline
SVD&        48  &  4\\
\hline
CDC&       75   &  8\\
\hline
TOP&        16   &  2\\
\hline
ARICH&    18   & 2 \\
\hline
ECL&        26    & 2 \\
\hline
KLM&       8   & 1  \\
\hline
TRG&   12     & 1 \\
\hline
\end{tabular}
\label{CPRvsPCIe40}
\end{table}

\subsection{PCIe40 module}
\begin{figure}[t]
\begin{subfigure}{0.31\textwidth}
    \includegraphics[width=3.5in]{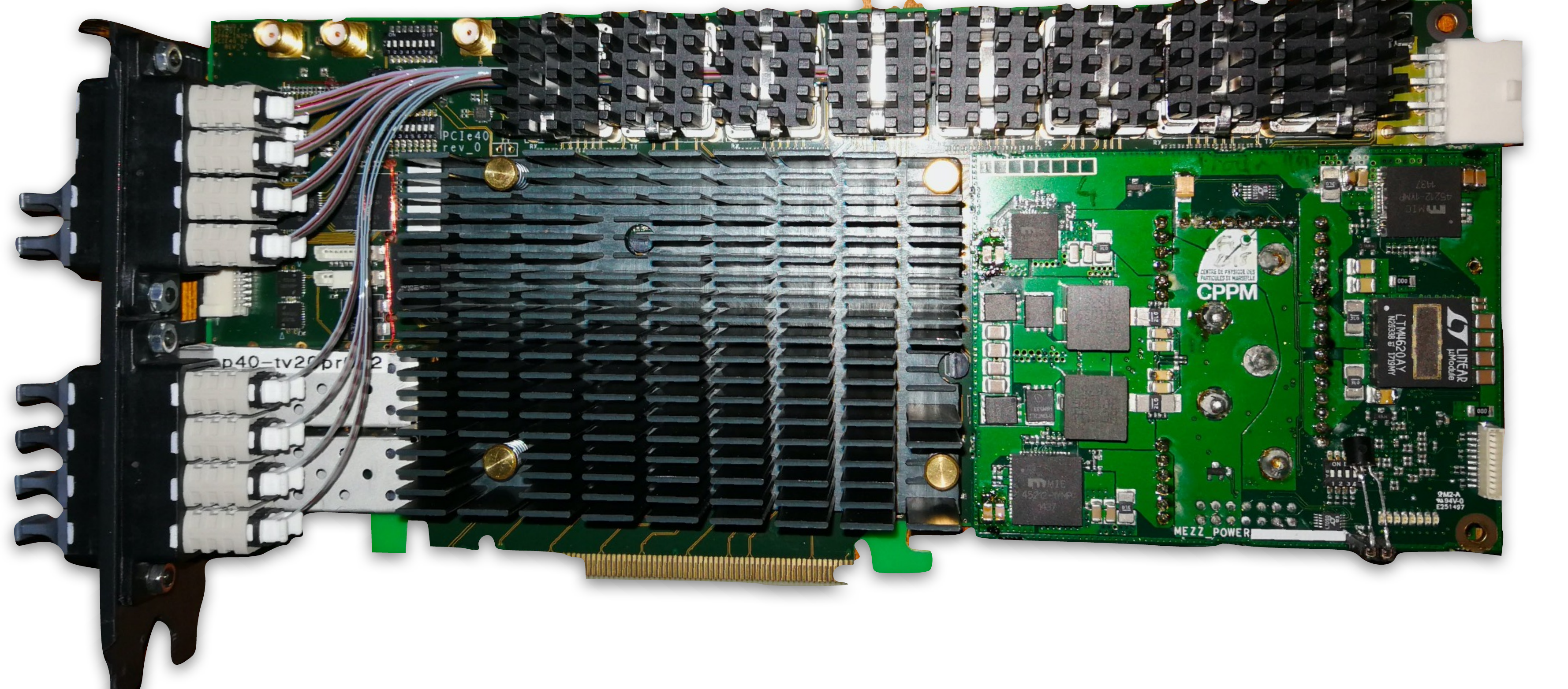}
    \caption{} \label{fig:pcie40a}
  \end{subfigure} \\%
  \begin{subfigure}{0.31\textwidth}
    \includegraphics[width=3.5in]{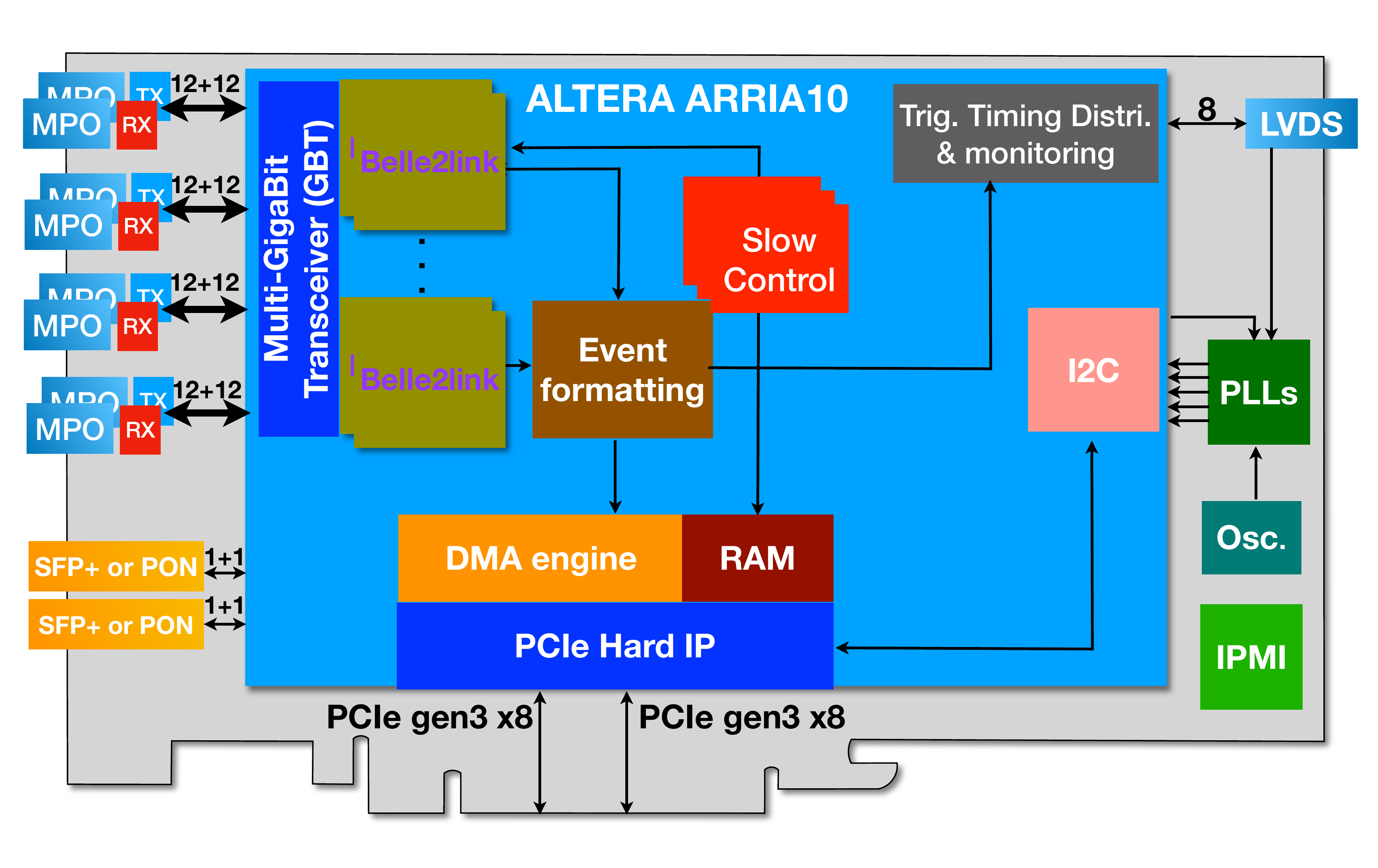}
    \caption{} \label{fig:pcie40b}
  \end{subfigure}
\caption{(a) Photo of PCIe40 board. (b) Block diagram of the PCIe40 board illustrating the key features of the board.}
\label{pcie40}
\end{figure}

The PCIe40 module is required to maintain interface compatibility with the currently running Belle II DAQ system.
As illustrated in Fig.~\ref{pcie40}, four pairs of onboard MiniPOD optical transceivers are used to receive or transmit data from or to the FEE through optical fibers. 
A high-density Intel Altera Arria 10 field-programmable gate array (FPGA) chip (1.15 million cells) provides the board with powerful reconfigurable logic capabilities.
Extra effort is required to implement the firmware of the Belle2link protocol on the Altera FPGA, as the current Belle2link logic is implemented on the Xilinx Virtex~5 FPGA. It has been verified that Belle2link is working well on PCIe40 by checking the cyclic redundancy check (CRC) errors during the data transmission.
PCIe40 is capable of running at up to 10 Gb/s for each link using the Gigabit Transceiver (GBT) architecture~\cite{GBT} with a maximum of 48 bidirectional links. 
In addition, there is an onboard LVDS connector with 8 links used as an interface to the Belle~II TTD system, and if it is necessary each board can be connected to  2 TTD modules (FTSW). 
Three external phase-locked loops (PLLs)  (Si5344 and Si5345) are used for clock jitter cleaning, and a clock source may be selected from the TTD system or from an onboard oscillator.
An embedded USB blaster on the front panel connected to the PC server enables remote downloading of the FPGA firmware. 
There is an over-heating protection system using the intelligent platform management interface (IPMI ) for the management. 
A point-to-point PCI-express Gen3$\times$16 links is used to connect the PC server and the PCIe40 board.
A PCIe40 board is presently installed on a 1U rack-mount server (Xeon E5-2640v4 10 Core/2.4GHz), which was prepared as a spare server for the current COPPER readout system. Multiple PCIe40 boards can be installed in a larger server. 

\subsection{Firmware development}

\begin{figure}[t]
\centerline{\includegraphics[width=3.5in]{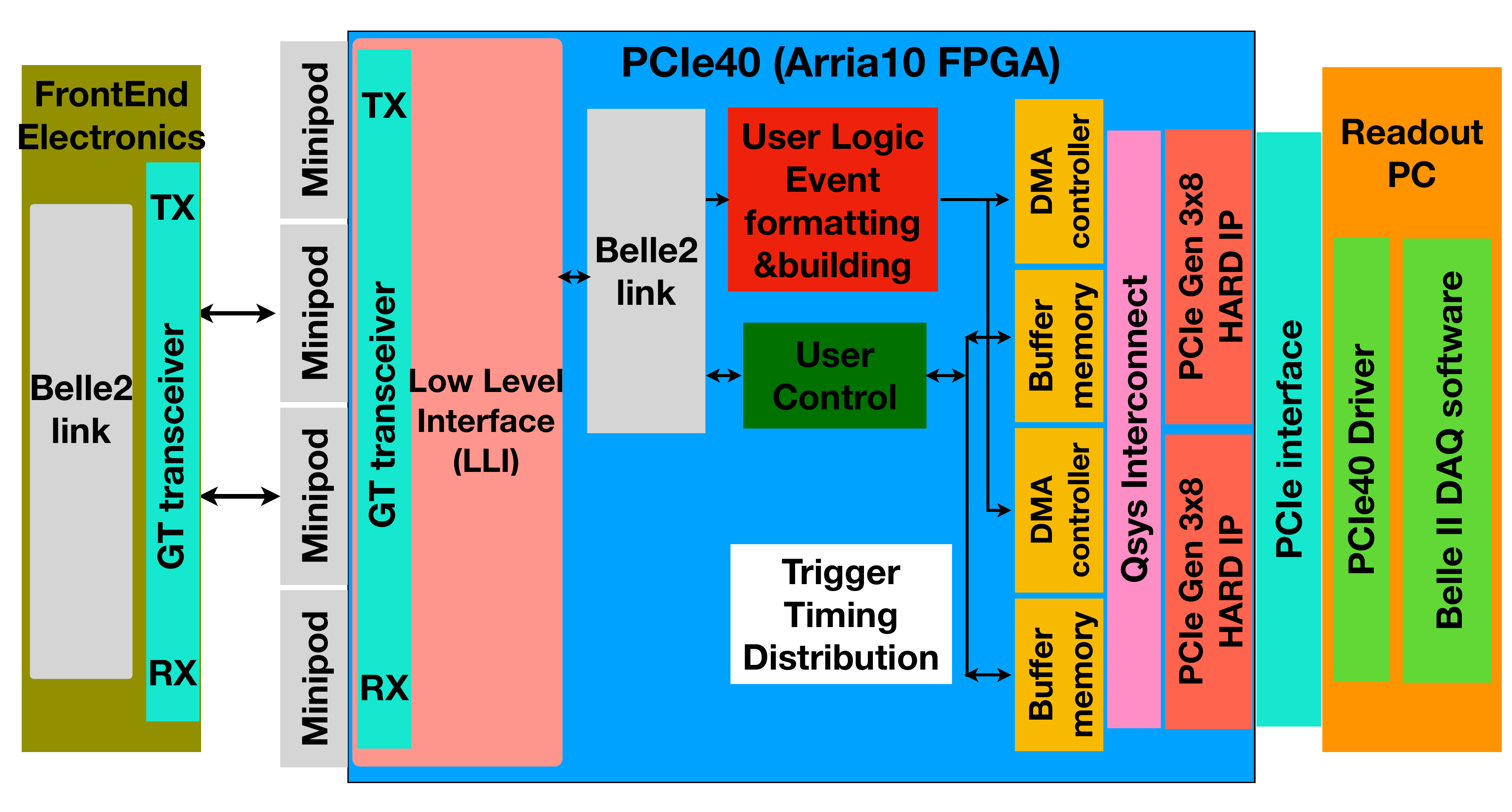}}
\caption{Block diagram of key features of firmware implemented on PCIe40. Front-end electronics and readout PCs are also shown to illustrate the data flow direction.}
\label{firmware}
\end{figure}

\begin{figure*}[h]
\centerline{\includegraphics[width=6.8in]{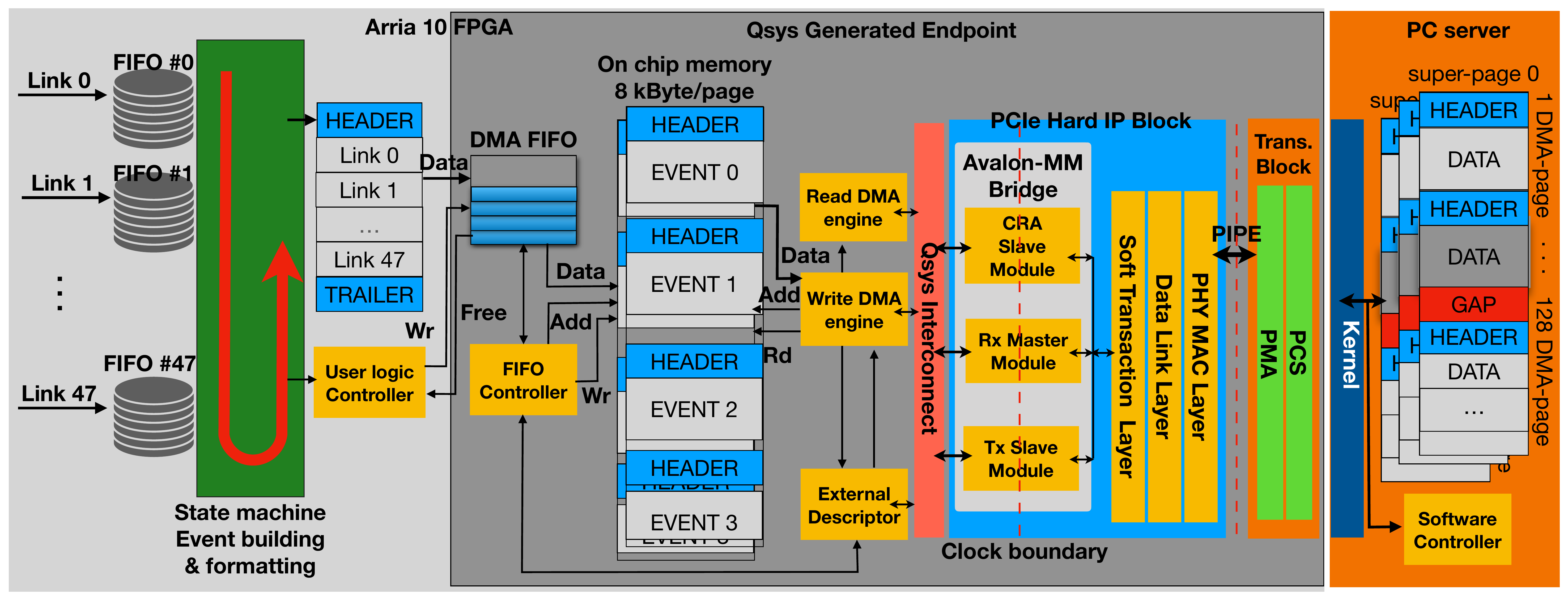}}
\caption{Block diagram of key features of firmware implemented on PCIe40 for data-flow processing and the software features for DMA. Firmware designs include two main functionality blocks, which are user logic part for event formatting, and speed-up event building and the PCI-express DMA core, which offers high bandwidth and low latency data transmission.}
\label{UL_DMA}
\end{figure*}   

The Belle2link protocol, b2tt protocol, and slow control functionalities which are implemented on the current COPPER-HSLB system, need to be integrated into the PCIe40 system. Event formatting and event building, as well as PCI-express hard IP-based direct memory access (DMA) are newly designed for PCIe40.
As shown in Fig.~\ref{firmware}, a generic firmware architecture is designed to provide a unified readout platform for all sub-detectors (except for PXD). 
A hardware abstraction layer, called low-level interface (LLI)~\cite{lli}, is designed for user control and initialization of the PCIe40 board.
The frontend side of the PCIe40 firmware includes the configuration of the GBT and decoding, synchronization, and aggregation of all GBT links.   
Meanwhile, it provides a simple interface for user application codes, such as the slow control unit. 
Thus, most user applications can connect with the LLI by simply retrieving from and pushing data to first-in-first-out (FIFO) interfaces.

The state machine of the Belle2link protocol is implemented on an Arria 10 FPGA. It includes both decoding and encoding of the event fragment data and slow control parameters based on the defined Belle2link format.   
Because the local bus widths of COPPER for address and data are 7-bit and 8-bit, respectively, the slow control feature was originally designed to access 8-bit registers mapped onto a 7-bit address space. This is known as the A7D8 access method for the Belle~II experiment. 
A16D32 and stream file features are upgraded based on A7D8 to satisfy different requirements of the sub-detectors. As listed in Table~\ref{slc}, each corresponding sub-detector uses one or two slow control access methods. All three features for slow control access are retained when upgraded to a new readout system. 
Address and data decoding and encoding for slow control accesses are moved from firmware to software. 
A FIFO is used as the interface to store addresses and parameters received or need to be transmitted by Belle2link. The FIFO is controlled by signals connected to the registers that are accessible using slow control software.
The ARICH detector uses the stream file method to transfer the firmware of the FEE, and KLM uses it to stream the threshold parameters for the RPC detector. 
To maintain a low error rate during transmission, the file to be streamed is split into packets of a certain length.
The data packet length is  100 words for ARICH and 6 words for KLM. 
This difference mainly depends on the FEE of ARICH and KLM.
  
\begin{table}
\caption{Access methods of each sub-detector for slow control}
\label{table}
\setlength{\tabcolsep}{3pt}
\begin{tabular}{|p{80pt}|p{40pt}|p{40pt}|p{55pt}|}
\hline
Sub-detector&  A7D8&  A16D32&  Stream file $^{\mathrm{a}}$\\
\hline
SVD&        $\checkmark$&  $\times$ & $\times$\\
\hline
CDC&       $\times$ & $\checkmark$&$\times$\\
\hline
TOP&       $\times$ & $\checkmark$&$\times$\\
\hline
ARICH&   $\times$ & $\checkmark$& $\checkmark$\\
\hline
ECL&       $\checkmark$ & $\checkmark$&$\times$\\
\hline
KLM&   $\times$ & $\checkmark$& $\checkmark$\\
\hline
\multicolumn{4}{p{245pt}}{$^{\mathrm{a}}$ Stream file method includes streaming a complete file, such as FEE firmware, and a defined parameter length.}
\end{tabular}
\label{slc}
\end{table}

Figure~\ref{UL_DMA} illustrates the key features of data processing on the PCIe40 board and readout PC server. 
A user logic block is designed to handle event building and data formatting, and a PCI-express hard IP-based DMA  architecture offers a fully integrated, flexible, and highly optimized solution for high bandwidth and low latency data transmission. 
With DMA, data are copied from the FPGA on-chip memory into the memory of the readout PC server without preempting the CPU.
The event fragment data are received by each Belle2link and buffered to a corresponding FIFO, and then the generic event formatting and event building logics are implemented to handle the data based on the firmware. 
The state machine is designed to merge fragments of the 48 links sequentially into an event and to remove redundant header and trailer information of each link, after it was checked.  
Subsequently, redefined event header and trailer information is added to the built event. 
The merged event data are then passed to the DMA data FIFO with a size of 32 kB, used as an interface to the PCIe DMA block.

The Arria 10 Avalon-MM (Memory Mapped) DMA interface~\cite{DMA} is used to easily design the PCIe protocol. As shown in Fig.~\ref{UL_DMA}, the design includes the following main features: on-chip memory,  PCIe Hard IP, and DMA descriptor controller. 
The built event data from the user logic block are stored in the on-chip memory having a 256-bit width and 8 kB pre-DMA page. 
Conversely, 128 DMA pages are contained in a DMA super page on the PC server side.
The DMA read and write operations are managed by an external DMA descriptor mechanism, which is separate from the DMA engine.
The procedure of sequentially copying data from the on-chip memory by reading out the PC server's memory is based on the descriptor table. The descriptor table contains the data source address (a pointer in the on-chip memory mapped fabric of the FPGA), destination PC memory's address (in the PCIe bus address space), and the transferred data size. 
The software controller first programs the descriptor controller's register with the location, size, and the number of the descriptor table; subsequently, the Read DMA engine issues a memory read request to forward the entire descriptor table to the FIFO controller.  
When the FIFO controller receives a list of free DMA super pages, the firmware DMA controller fills these DMA super pages without further software intervention.
The controller then fetches the table and requests the Write DMA engine to transfer the data from the Avalon-MM to the PCIe domains with one descriptor at a time.
When the data transmission is completed, the controller sends the DMA status upstream.

An optimized handshake among the user logic block, PCIe-based DMA, and the corresponding software driver is essential. A certain large buffer capability, careful synchronization within the DMA controller, and sufficient data processing power of the readout PC server ensure that data can be simultaneously transferred downstream at the highest possible rate with efficient utilization of resources. 

The TTD system is designed for trigger and timing distribution and busy handshake control for the Belle~II experiment. 
A unified bidirectional communication based on a 254 Mb/s serial data link, called b2tt, is used as the interface between the PCIe40 board and the TTD system. 
A TTD interface with one b2tt link is implemented for the PCIe40 board to handle all 48 belle2links simultaneously. 
Currently, architecture to handle 48 belle2links individually is being developed. 
Thus, when a link has an error or is busy, the corresponding issue can be successfully collected by the TTD system.
A 127-MHz external clock derived from the SuperKEKB accelerator's radio frequency (RF) is distributed to the PCIe40 board and used as the system clock for the firmware.

\subsection{Software development}
Data readout and slow control access based PCI-express interface are new features compared to the COPPER-HSLB system.
The software driver and library are developed based on C/C++, and they mainly provide the functionality for accessing and controlling the data acquisition or configuration of the PCIe40 board. On the PCIe40 side, the base address register 0 (BAR0) and BAR2 of PCI-express are designed to be used for slow control access and DMA access, respectively. The corresponding driver and library were developed individually. 
The driver is involved in managing physical memory, and the library provides utilities for accessing registers, performing tests, etc. 
The driver for DMA is also used to manage the DMA access and provide the functionalities of accessing the availability status of memory for DMA, checking the arrived data, and then applying operations to the arrived data. As shown in Fig.~\ref{UL_DMA}, the unit used for individual DMA transfer is the DMA page, which is 8 kB per page, and this size depends on the Linux constraint. On the PC servera-side, 128 DMA pages are contained in a DMA super page of 1 MB. A typically large buffer of several GBs is allocated for each DMA channel. The management of the DMA access and memory allocation is based on the super page. 

\begin{figure}[t]
\begin{subfigure}{0.31\textwidth}
    \includegraphics[width=3.5in]{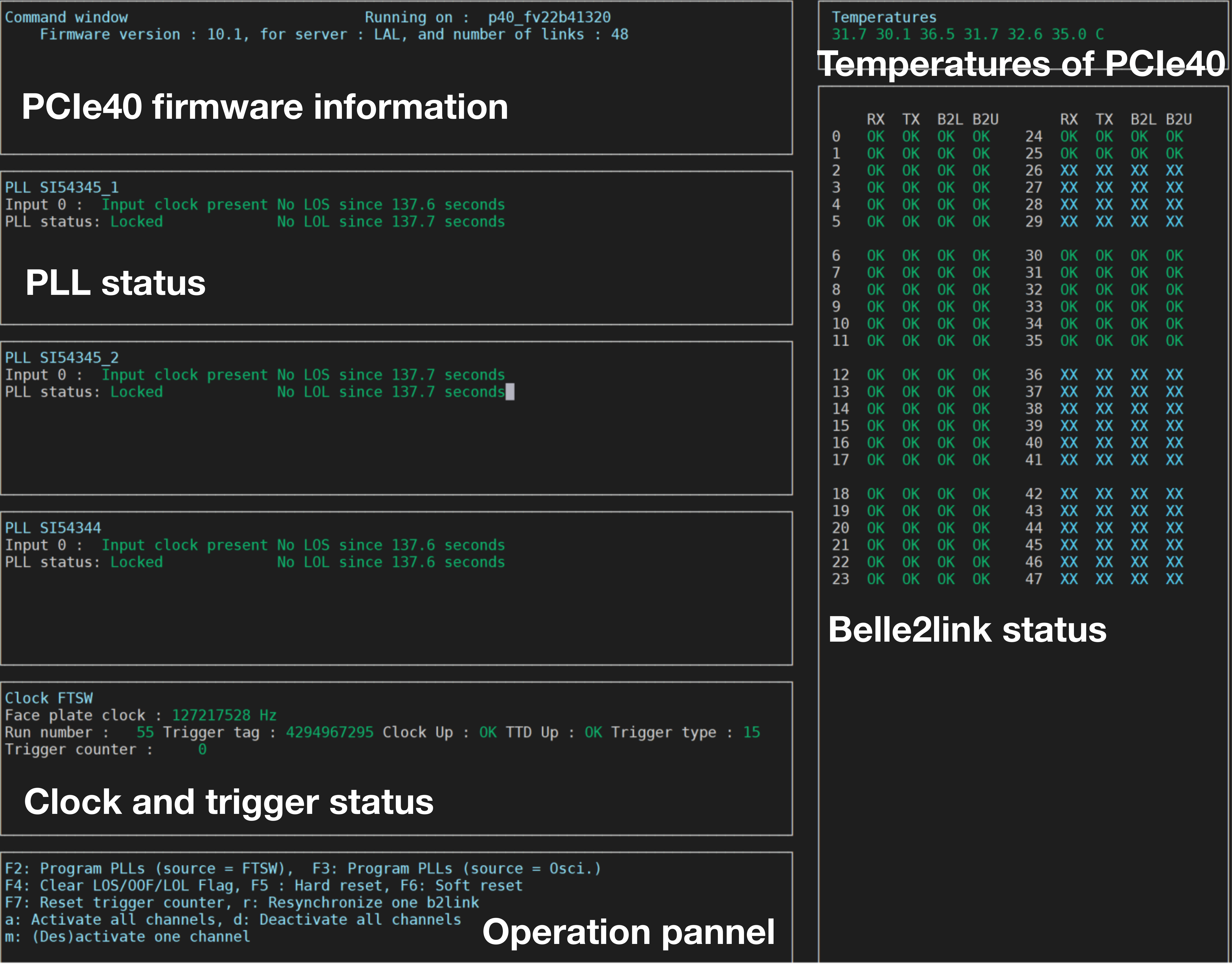}
    \caption{} \label{fig:monitora}
  \end{subfigure} \\%
  \begin{subfigure}{0.31\textwidth}
    \includegraphics[width=3.5in]{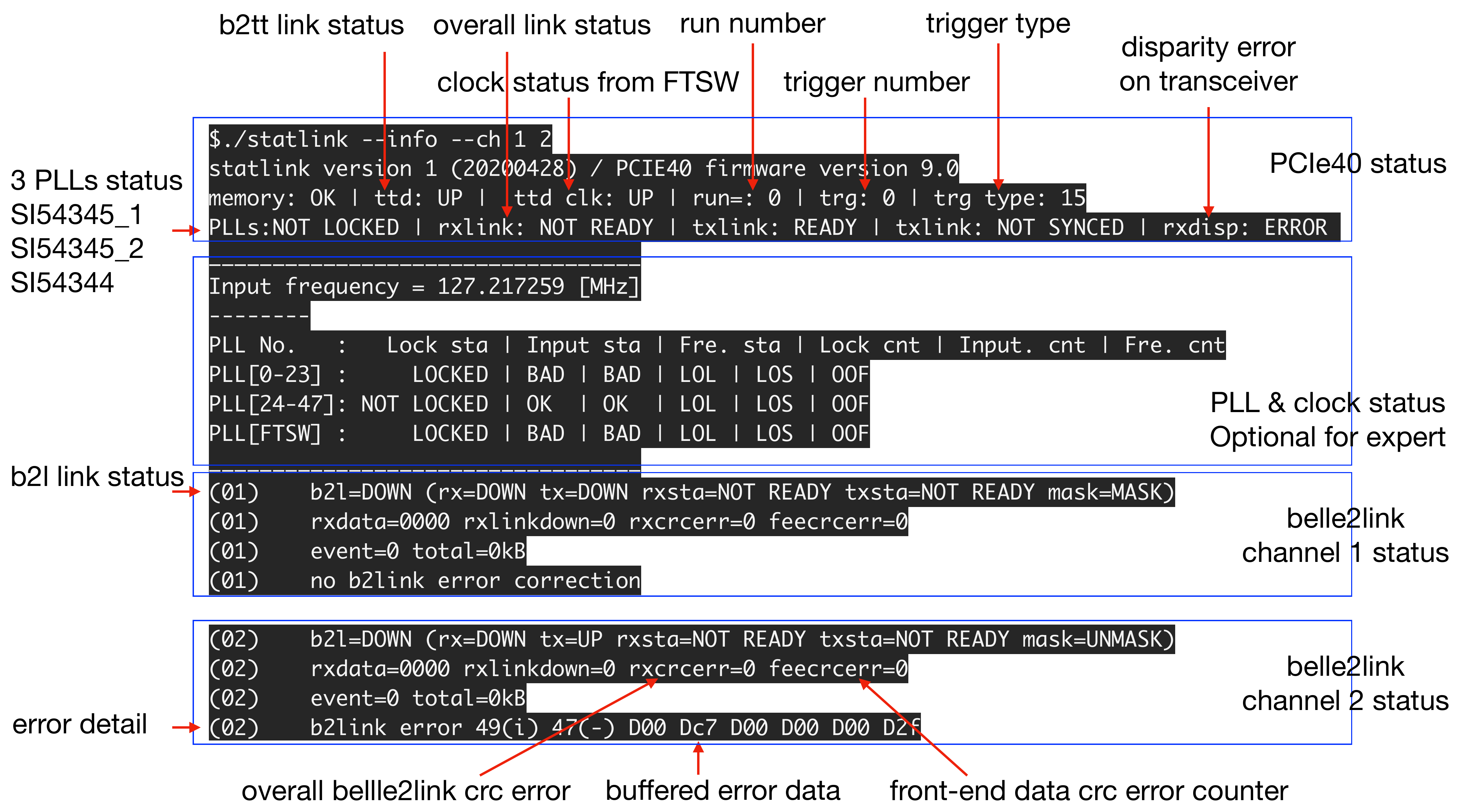}
    \caption{} \label{fig:monitorb}
  \end{subfigure}
\caption{(a) GUI view of a general  monitoring system for PCIe40 board and Belle2link. (b) Output of the monitoring system with specific error details. It can be integrated into the Belle~II log system.}
\label{monitor}
\end{figure}

An application layer, developed based on low-level drivers and libraries, contains individual utilities for high-rate data acquisition, slow control register access, monitoring, and configuration of the PCIe40 board, Belle2links, etc. 
An example of an application is shown in Fig.~\ref{fig:monitora}. A Python script is designed to show a GUI window with firmware information, PLL status, clock and trigger information, and all Belle2link status. All this information is used for a monitoring purpose, it can help the operator to quickly investigate the readout system if an issue occourred.  An operation panel provides commonly used configuration options to rapidly switch the clock source, resynchronize the Belle2link for link establishment, and mask or unmask an individual or all Belle2links. 

\subsection{Software development for integration with Belle~II DAQ system}
There are mainly two blocks of software that need to be upgraded for PCIe40 board in order to integrate with the current Belle~II DAQ system. 
One is related to the slow control system, and the other to the data readout system. 
The slow control feature of the current COPPER system was designed based on the COPPER local bus. 
Similar features must be developed for the PCIe40 board. 
As shown in Fig.~\ref{fig:monitorb}, a commonly used tool for investigating and debugging the cause of an error occurring in the readout system has been developed. 
It shows inclusive PCIe40 status, individual link status, counters of CRC errors, and buffered corrupted data.
Meanwhile, this output format can be saved into the logging system thus can be easily integrated into the current Belle~II logging and monitoring system.

\begin{figure}[t]
\centerline{\includegraphics[width=3.5in]{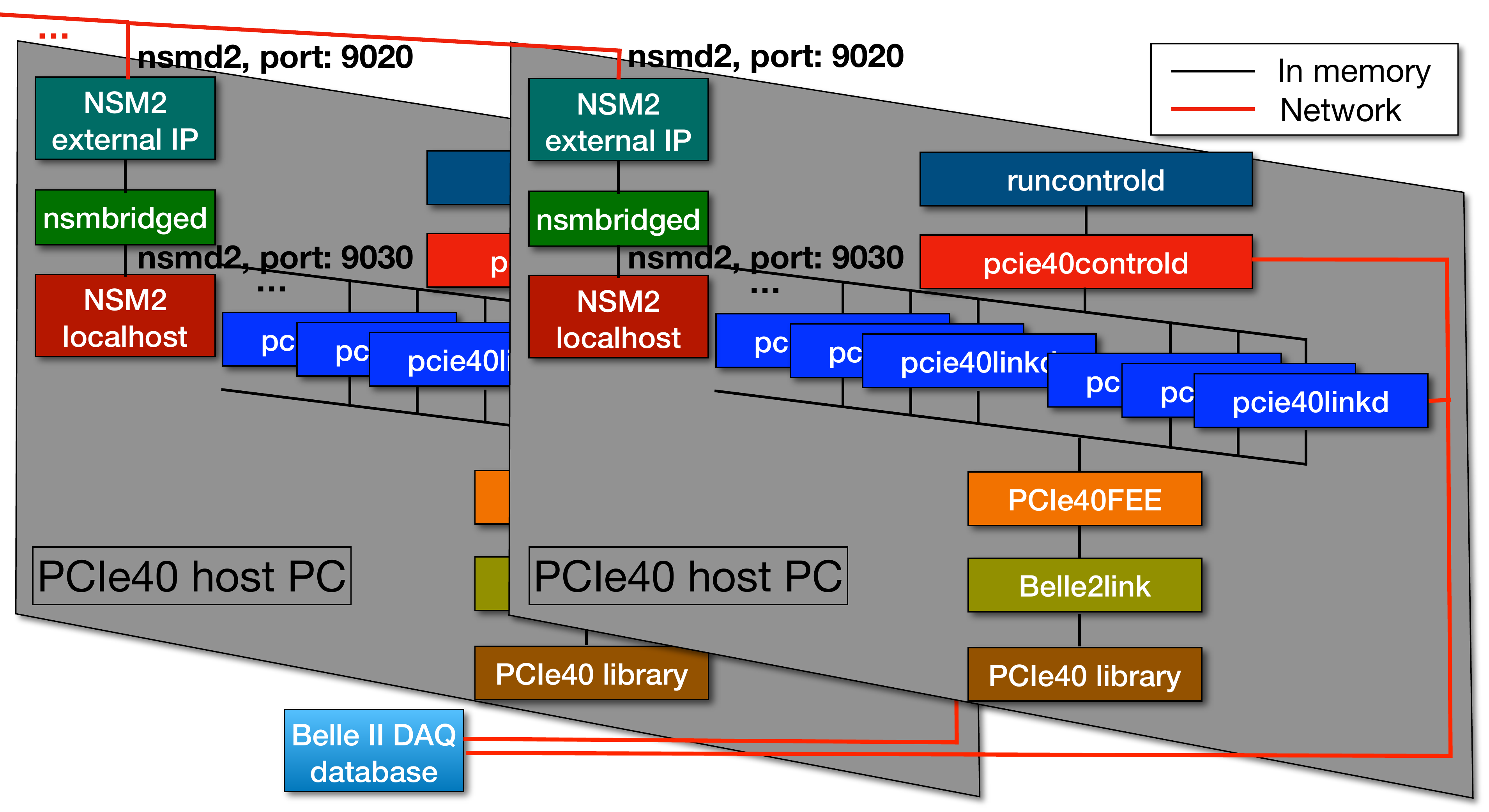}}
\caption{Block diagram of slow control architecture on readout PC; a NSM2 based system, involving several newly developed daemon processes, such as pcie40controld and pcie40linkd, has been developed. }
\label{NSM2}
\end{figure}

The Belle~II DAQ slow control system is designed based on a network interface called network shared memory 2 (NSM2)~\cite{nsmd2}. 
The application nodes are C/C++ daemon programs, which are used to manage the slow control functionalities. 
Figure~\ref{NSM2} illustrates the slow control architecture on the PCIe40 readout PC with newly developed daemon processes, such as pcie40controld and pcie40linkd. 
pcie40linkd is used to manage and control an individual Belle2link, and pcie40controld is designed to manage and control all pcie40linkd processes running on the readout PC. 
PCIe40FEE is a sub-detector-specified library, where the sub-detectors can define their functionalities to control their detectors through Belle2link. 

The data readout related modifications to be integrated with the current DAQ system are relatively small.
Data readout software with the new DMA scheme has already been developed. The data format of the merged event is slightly different and needs changes for data compression. Because the data from 48 links are processed in one readout PC, the upgrade of the related software from 4 links for COPPER to 48 links for PCIe40 is essential.

\section{Performance of new readout system}
\label{sec:performance}
Firmware and software for the new readout system have already been developed. As discussed in Sec.~\ref{sec:requirement}, it is necessary to confirm that the new readout system fulfills the requirements of the Belle~II DAQ system. A test bench was constructed for the demonstration of the readout system. Meanwhile, another readout system was built and connected with the Belle~II TOP and KLM sub-detectors at the Electronics-Hut in Tsukuba Hall, at KEK, which is used to host the electronics of the Belle~II experiment. Slow control and data readout tests were performed using the TOP and KLM on-site detectors.

\subsection{Test bench}
\begin{figure}[t]
\centerline{\includegraphics[width=3.5in]{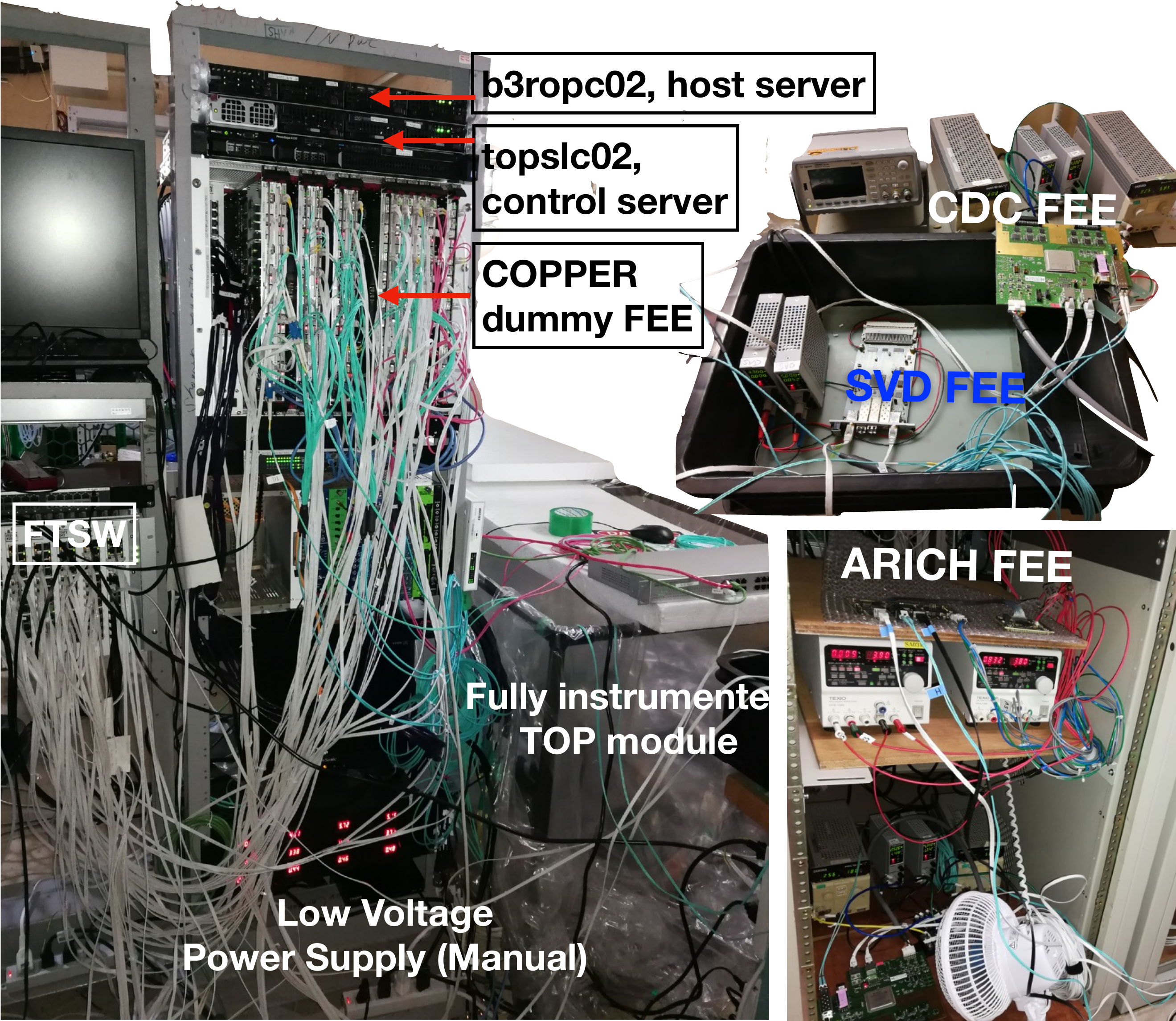}}
\caption{Overview image of test bench for DAQ readout upgrade. FEEs of TOP, CDC, ARICH, SVD, and KLM and  46 dummy FEEs based on COPPERs were connected to the PCIe40 board for the performance test.}
\label{B4TB}
\end{figure}

As shown in Fig.~\ref{B4TB}, we constructed a test bench for the readout system, specifically involving the FEEs of the CDC, SVD, TOP, and ARICH and 46 dummy FEEs based on COPPER-HSLB, for demonstration.
The PCIe40 board was installed on a host PC server, and the electric power was provided by the host PC. 
The firmware was downloaded using an embedded USB blaster on the PC. 
Forty-eight FEEs were connected to the PCIe40 board using 8 Multiple-Fiber Push-on/Pull-off (MPO) fibers, and each channel could be activated or deactivated by controlling the corresponding register on the PCIe40. 
The clock and triggers were distributed to PCIe40 and FEEs by the TTD system. The two types of triggers are evenly spaced pulse triggers and random triggers with intervals of pseudo-Poisson distribution, which can be issued by the TTD system with an adjustable input trigger rate.

\subsection{Performance of data readout}
\begin{figure}[b]
\centerline{\includegraphics[width=3.5in]{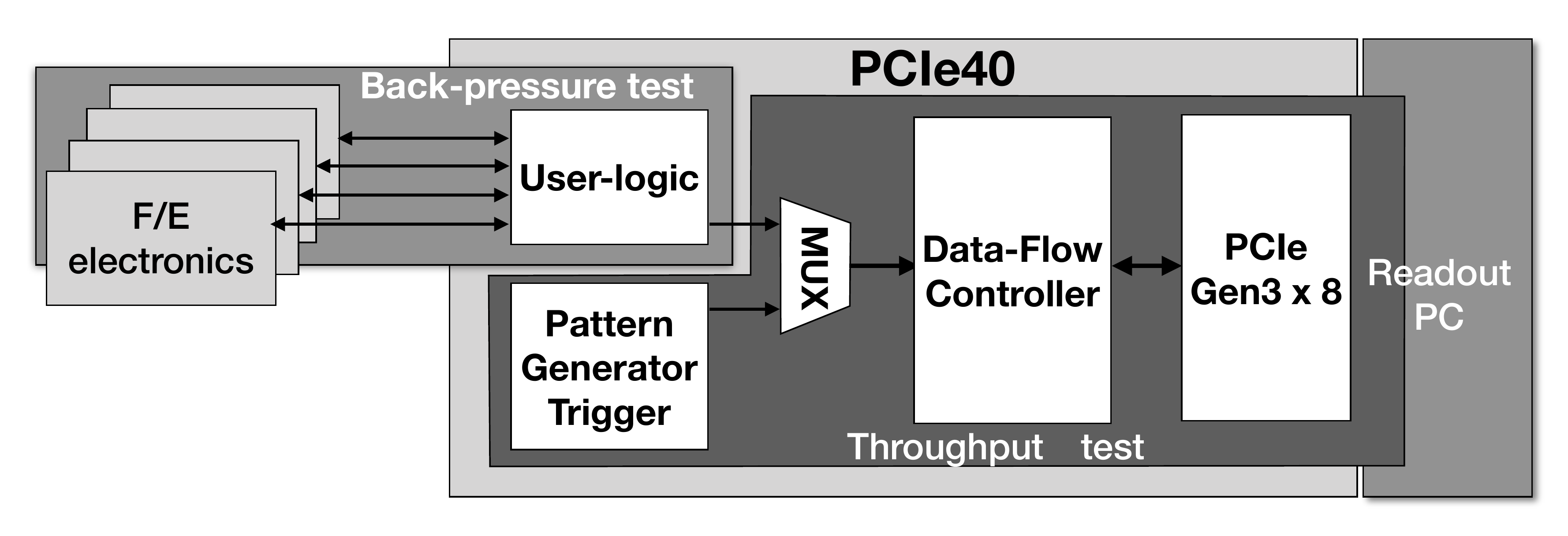}}
\caption{Schematic view of components (throughput-test block and back-pressure-test block) involved in the two performance tests.}
\label{readout_test}
\end{figure}

Firstly, we measured the throughput of PCIe DMA data-transfer by using a throughput test block shown in Fig. ~\ref{readout_test}, where a pattern generator was implemented as the data source instead of data from the FEEs.
A FIFO was added for back-pressure monitoring. 
Measurements at high event-rates were performed with a pulse trigger control to confirm the capability of the new readout hardware.
\begin{enumerate}
\item The back-pressure was observed in 10\% of events when using a pulse trigger with 470 kHz trigger rate and an event size of 8 kB, for which the corresponding data transfer rate was 31 Gb/s, 
\item No event was lost if the trigger rate was reduced to 260 kHz and when the data transfer rate was 17 Gb/s.
\end{enumerate}
It is reported that the performance of a single instance of the DMA controller bonded to a PCIe$\times$8 Gen3 interface can consistently reach 50 Gb/s~\cite{pcie40_100G}, and it can eventually be increased to 100 Gb/s using the PCIe$\times$16 Gen3 interface completely.
Conversely, the current performance of the data readout system already fulfills the Belle~II DAQ upgrade requirements, while the throughput limit of the new readout system is 10 Gb/s, which is determined from the new readout PC.

Secondly, we demonstrated a back-pressure scheme in data-transfer from 46 COPPER-HSLB dummy FEEs as data sources to a PCIe40 board, by using a back-pressure test block in the FPGA of the board shown in Fig.~\ref{readout_test}. 
No back-pressure or lost events were observed with the new readout system under a 43 kHz trigger rate and 1 kB~/~event~/~link data readout. Here, the trigger control was set with a restriction of no more than 10 triggers within 130 $\mu$s, which was determined from the SVD FEEs. 
Meanwhile, 2.5\% of the events issued a back-pressure under a 26-kHz trigger rate when a 0.8~ms latency was artificially inserted into the FEE data source.   
The back-pressure is due to the limited size of the FIFOs, which are implemented to buffer the data from the FEEs for event building.   
The on-chip memory size of the PCIe40 board is only $\sim$ 70 Mbits. As the memory of PC server is much larger, this issue will be solved by transferring the event building from the FPGA logic to the software in the future.    

\subsection{Performance of slow control system}
All three slow control access methods using the PCIe interface were tested on the test bench. The single access speed of the A16D32 method was tested by reading the register of the CDC FEE 1$\times 10^6$ times. The average access time was 83 $\mu$s/access, while the COPPER-HSLB system took $\sim$1 ms. The streaming file method was tested using the FEEs of KLM and ARICH. The streaming speed was $\sim$360 kB/s for KLM, which is comparable with the COPPER-HSLB system ($\sim$350 kB/s). Both the PCIe40 and the COPPER-HSLB system took less than 2 s to transfer the 3 MB ARICH firmware. 
A new feature for accessing different individual links in parallel is available for the PCIe40 board, which sufficiently reduces the time required for configuring the FEEs of a sub-detector, 
while, for the COPPER-HSLB system, the four links need to be accessed sequentially.

\begin{figure}[h]
\centerline{\includegraphics[width=3.5in]{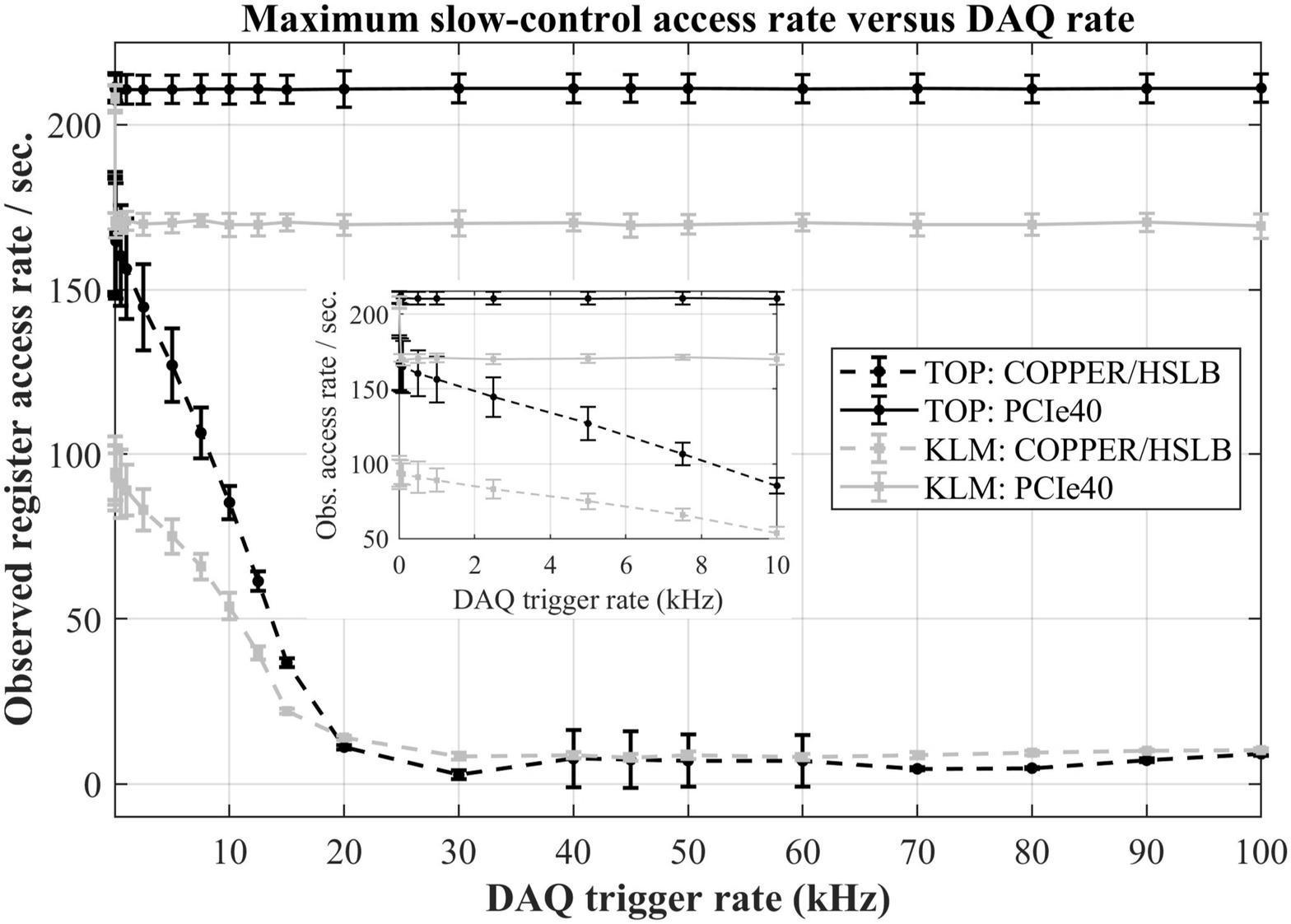}}
\caption{Register accessing rate when simultaneous data acquisition is performed with different trigger rates. The curve showing a decreasing accessing rate from 100 to 0 corresponds to the access of a KLM FEE register using the COPPER-HSLB system, while the accessing rate retains a constant value of 170 for the PCIe40 board. A similar tendency is observed for TOP in which the accessing rate decreases for the COPPER-HSLB system when the DAQ trigger rate increases and remains constant for the PCIe40 board.}
\label{slc_perf}
\end{figure}

A register access test was performed together with data acquisition under different trigger rates. As shown in Fig.~\ref{slc_perf}, the register accessing rates for TOP and KLM based on the COPPER-HSLB system decrease from 170 and 100 to 0, respectively, when the data acquisition trigger rate increase from 0 to 30 kHz.
This indicates that there is a limitation of bandwidth for slow control and data acquisition for the COPPER-HSLB system. 
However, the corresponding register accessing rates for TOP and KLM based on the PCIe40 board retain a constant value even when the data acquisition trigger rate is increasing. There was no limitation of bandwidth observed for the PCIe40 board.     

Another complete slow control test was performed  with the on-site FEEs of TOP and KLM. All 64 FEEs of TOP were connected to 2 PCIe40 boards that were hosted by 2 separate servers, and 32 FEEs of KLM were connected to a single PCIe40 board. The slow control test was performed based on the upgraded software architecture, as shown in Fig.~\ref{NSM2}.
The speeds of configuring FEEs and setting the threshold for both TOP and KLM detectors based on the PCIe40 board are compatible with the COPPER-HSLB board.  
The configuration for TOP was confirmed by taking a short run and checking the data quality.
The threshold for KLM was verified by checking the output raw data file size by varying the threshold value. 
It is confirmed that the new software architecture for slow control works properly. 

\section{Preparation and replacement}
\label{sec:replace}
The mass production of 31 PCIe40 boards for Belle~II DAQ upgrade is completed, and the boards have arrived at KEK after they have been successfully tested at CERN.   
Three boards were installed on 3 servers for the readout system replacement of TOP and KLM. 
MPO fibers are connected to the patch panels located close to the COPPER-HSLB system of TOP and KLM.  
A 10 Gb/s network interface card (NIC) is installed on host PC servers for downstream connection. 
A 40 Gb network switch is installed close to the readout PCs for connection with the network switch located in the HLT server room.
The replacement of the readout system for TOP and KLM is scheduled for summer 2021, and the replacement for other sub-detectors will take place afterwards. 
The current COPPER system will stand by for some time after the replacement. In the case of a serious problem, it can be rolled back to the COPPER system quickly.

\section{Conclusion}
\label{sec:conclusion}
A proposal based on the PCIe40 board was adopted for the Belle~II DAQ upgrade.
It is necessary to retain the main features of the current readout system and reduce the modifications required for the sub-detectors when upgrading to the new readout system.
The developments of FPGA firmware and software are almost complete.
The PCI-express hard IP-based DMA architecture of the new readout hardware is capable of handling a 260-kHz trigger rate with 17 Gb/s data throughput without data loss.
No back-pressure or event loss was observed in data-transfer from 46 dummy FEEs to the PCIe40 board under 43-kHz trigger rate and 1 kB/event/link data size.
Three accessing methods used for slow control offered similar or better performance with respect to the current COPPER-HSLB system. 
A new feature that enables accessing individual links in parallel improves the configuration speed of the sub-detectors.
It has been confirmed that the new software architecture for slow control works properly.
 The replacement of the Belle~II readout system is planned to start from TOP and KLM sub-detectors for summer 2021, and the replacement of the COPPER modules of the other sub-detectors will be done afterwards.

\end{document}